**Letter to the Editor and Comments on: Intravital dynamic and correlative imaging reveals diffusion-dominated canalicular and flow-augmented ductular bile flux**

**Lutz Brusch[1], Yannis Kalaidzidis[2,6], Kirstin Meyer[3], Ivo F. Sbalzarini[2,4,5,6] and Marino Zerial[2,5,6]**

1 Center for Information Services and High Performance Computing, Technische Universität Dresden, 01062 Dresden, Saxony, Germany
2) Max Planck Institute of Molecular Cell Biology and Genetics, 01307 Dresden, Saxony, Germany
3) UCSF, Cardiovascular Research Institute, Department of Biochemistry and Biophysics, California, USA
4) Chair of Scientific Computing for Systems Biology, Faculty of Computer Science, Technische Universität Dresden, 01187 Dresden, Saxony, Germany
5) Center for Systems Biology Dresden, 01307 Dresden, Saxony, Germany
6) Cluster of Excellence Physics of Life, Technische Universität Dresden, 01187 Dresden, Saxony, Germany

Bile, the central metabolic product of the liver, is secreted by hepatocytes into bile canaliculi (BC), tubular subcellular structures of 0.5-2 µm diameter which are formed by the apical membranes of juxtaposed hepatocytes. BC interconnect to build a highly ramified 3D network that collects and transports bile towards larger interlobular bile ducts (IBD). The transport mechanism of bile is of fundamental interest and the current text-book model of „osmotic canalicular bile flow", i.e. enforced by osmotic water influx, has recently been quantitatively verified based on flux measurements and BC contractility (Meyer et al., 2017) but challenged in the article entitled "Intravital dynamic and correlative imaging reveals diffusion-dominated canalicular and flow-augmented ductular bile flux" (Vartak et al., 2020). We feel compelled to share a series of arguments that question the key conclusions of this study (Vartak et al., 2020), particularly because it risks to be misinterpreted and misleading for the field in view of potential clinical applications. We summarized the arguments in the following 8 points.

1. <u>A curb of the experimental approach lies in the diffraction limit of light microscopy.</u>
The authors of (Vartak et al., 2020) measured bile flux kinetics in BC vs. IBD using a Fluorescence Loss After Photoactivation (FLAP) approach based on photoactivatable bile tracers which become fluorescent upon focused illumination. The estimation of fluorescence signal upon local photo-activation over time is at the basis of the conclusion that diffusion dominates over advection (flow) in bile flux through BC. This conclusion however depends on the assumption that the tracer can be photoconverted mainly, if not exclusively, in the BC in the focal plane of the microscope. However, given that the point spread function of the microscope is markedly elongated along the z-axis (see Fig. 1), it is impossible that the photoactivation achieved using the 405nm laser can be confined within the focal plane of the scanning microscope. This means that a significant fraction of fluorescence tracer must be photoactivated in the surrounding space (± 10 µm), i.e. the cytoplasm of hepatocytes, and contributes to the signal detected in the BC. Even if photoactivation occurs in an area where fluorescence may be suppressed by the pinhole of the scanning microscope during image acquisition, in a short time the fluorescent tracers will move into the focal plane and become visible. As consequence, along with the transport of the tracer there will also be continuous secretion of activated tracer from its cytoplasmic reservoir into the BC at the location of initial photoactivation, making it impossible to interpret the results unequivocally. In contrast, the measurements of the blood vessels are less sensitive to such artefacts because a) the diameter of sinusoids (~6 µm) is closer to the z-diffraction limit; b) physiologically, there is no significant back-flux of marker from hepatocyte to blood. This fundamental difference implies that the blood measurements are not appropriate controls for this experimental system. The same argument applies to the IBD measurements which show the expected advective flux.

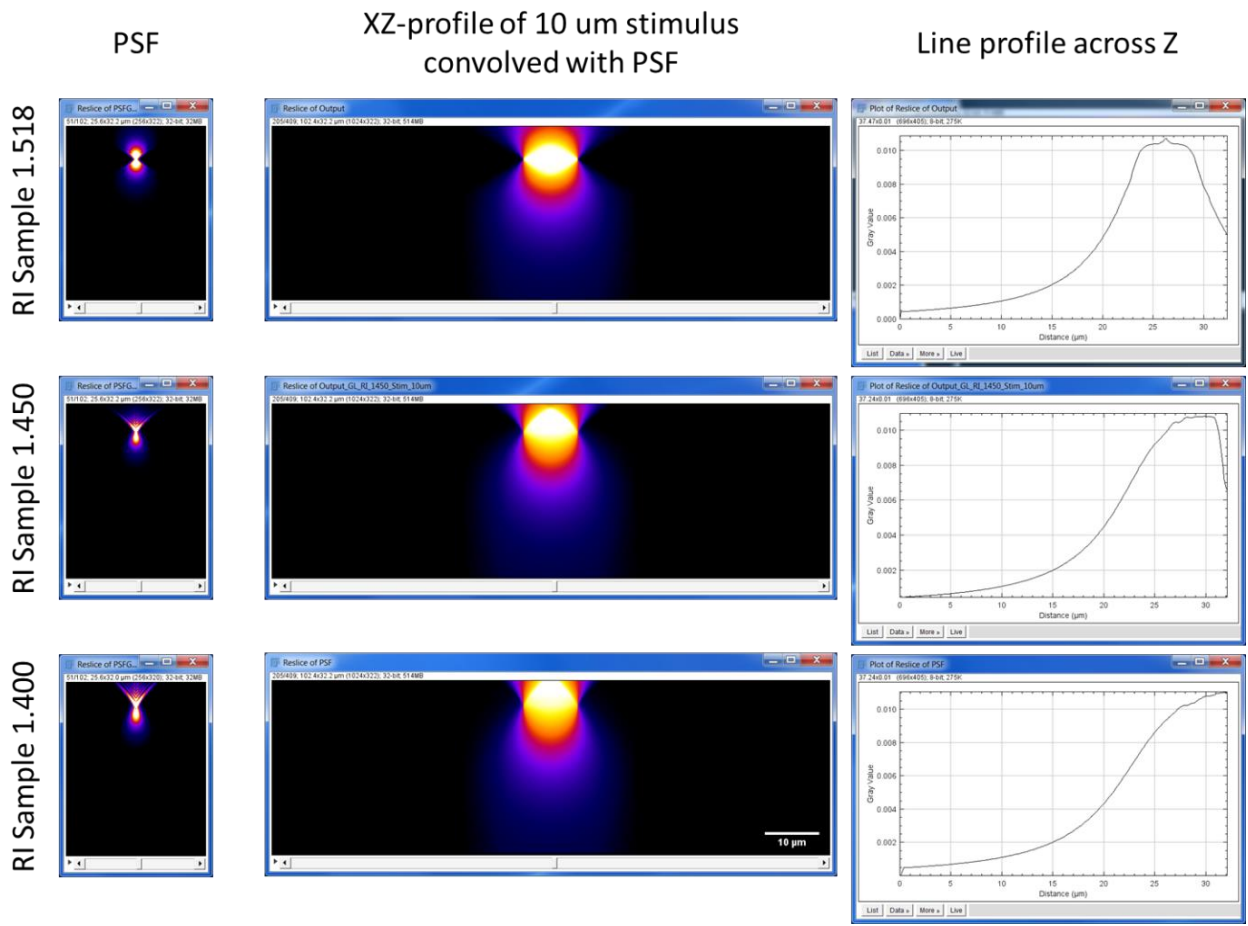

**Figure 1. Theoretical photo-activation intensity distribution** calculated by the PSF calculator (Image J) for different refraction indexes of the tissue, as indicated. The left panels present the point spread functions, the middle panels the intensity distributions in the xz section of the uncaged volume and the right panels show the graphs of intensity in the z-direction. Note that the characteristic size of uncaged volume in the z-axis is ~ 20 µm. This is roughly the size of hepatocytes, indicating that the photoactivation is not restricted to the BC but can include the whole cytoplasm of adjacent hepatocytes. The parameters for calculation are presented in Fig. 2.

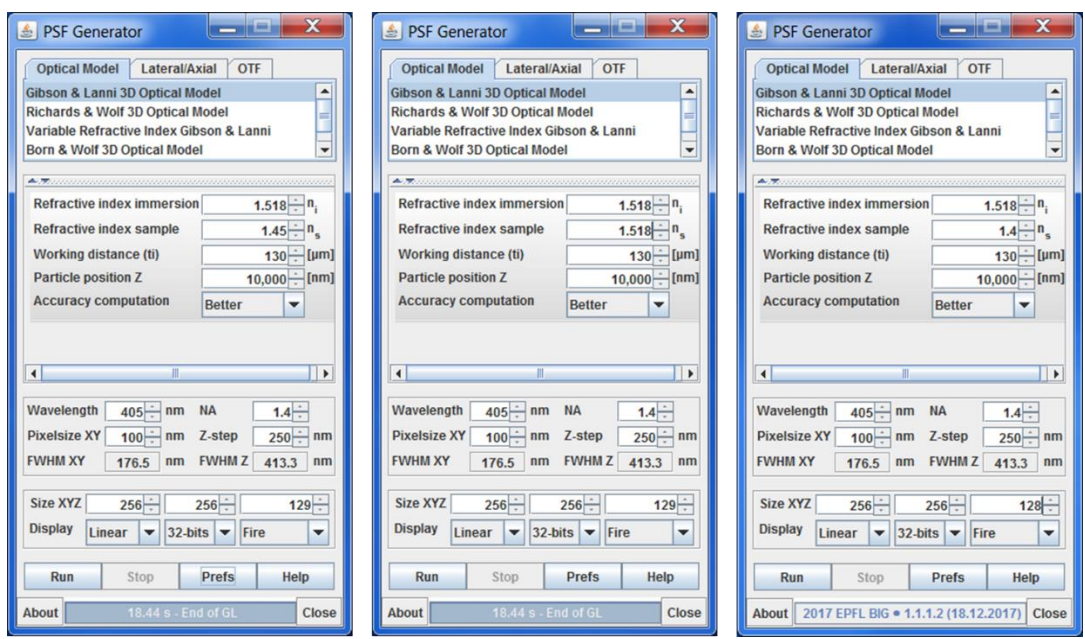

**Figure 2. Parameters** used for the PSF calculation for the uncaged volume (Fig. 1).

2. <u>BC are not empty tubes.</u> Another drawback of the FLAP setting is the fact that the volume of microvilli is substantial, accounting for up to ½ the volume of the BC lumen (Pfeifer and Rous, 1980; Meyer et al., 2017). This implies that fluorophores within the cytoplasm of microvilli or in the sub-apical cytoplasmic volume cannot be advected (as they are not in the BC lumen). These cytoplasmic signal reservoirs and the ongoing local release of activated tracer at the initial position will artificially maintain the position of the concentration maximum (centroid), which can be misinterpreted as absence of flow.

3. <u>Is there really no flow?</u> The data of Fig.1C (middle row) in (Vartak et al., 2020) do show an asymmetry in the movement of the photo-activated tracer (faster up-ward than down-ward and right-ward), on a distance ~10-20 µm over 20 sec, which is absolutely in line with previously published values of bile flow in BC (Meyer et al., 2017). It is unclear by which algorithm the linear "distance"-axis of the quantitative intensity profile for BC has been chosen on the imaged plane, as different conclusions are expected when choosing the "distance"-axis for instance horizontal versus vertical in the representative images shown.

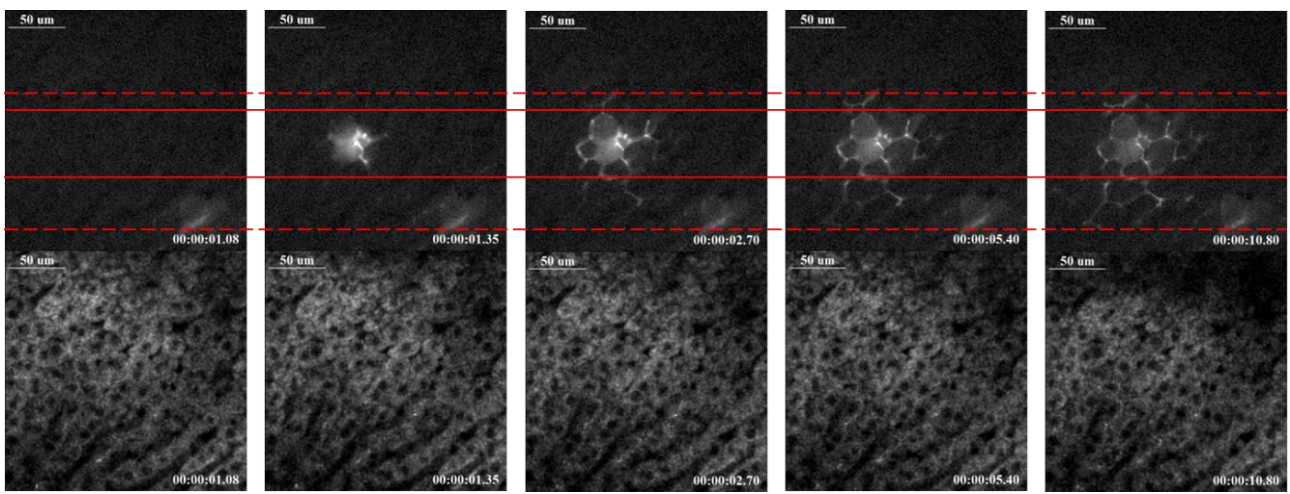

**Figure 3. Data analysis.** Selected frames from Suppl. Materials of (Vartak et al., 2020) (/photoactivation/basal/set(5)/25_mz.czi) demonstrate directional flux of the tracers in the middle zone of the lobule. The solid red lines denote top and bottom boundaries of the fluorescence signal upon photoactivation, the dotted lines denote the boundaries after 9.5 seconds of tracer propagation.

One has to give credit to the authors for sharing the raw microscopy data in the Suppl. Materials of (Vartak et al., 2020). These published raw data of photoactivation in the middle zone are presented here in Fig. 3. From the images, one can estimate that the tracer does not propagate symmetrically (from the solid to dotted lines in Fig. 3). From the differences between tracer propagation in the opposite directions, one can make a rough estimation of advection speed v~3 µm/sec, in line with previously reported data (Meyer et al., 2017), together with a higher diffusion coefficient than reported in (Vartak et al., 2020). To quantitatively compare the respective contributions of advection and diffusion to transport, the Péclet number Pe=L*v/D has to be calculated for the liver lobule and compared against 1 (see Appendix). For these measured quantities in the order of L~200 µm, v~3 µm/sec and considering D~3 µm$^2$/sec as done in (Vartak et al., 2020), the Péclet number is Pe=L*v/D~200, which is Pe >> 1 and therefore, transport is advection-dominated, and <u>not</u> diffusion-dominated.

4. <u>Is diffusion correctly estimated?</u> In the Materials and Methods section of (Vartak et al., 2020), the authors report the use of an incorrect analytical approach. The formula for the determination of half-life, which is used for the estimation of diffusion coefficients, has no physical justification because 1) the decrease in intensity of the marker from the center of propagation follows a hyperbola <u>instead</u> of an exponential function of time, 2) bleaching is <u>not</u> described by an additive linear term in the time course of intensity (- a * t), but by a multiplicative factor with negative exponent of time (* exp(-t/t0)), 3) the authors of (Vartak et al., 2020) used non-critically the formula for diffusion estimation of 2D FRAP-experiments rather than a formula for 3D, and 4) formulas for homogeneous open space neglecting the geometry of the diffusion space (which is a much more complex network of tubules). Altogether, these errors lead to an incorrect (i.e., under-) estimation of diffusion coefficients and advection.

5. <u>Is IVARICS applicable on intra-vital microscopy?</u> The authors of (Vartak et al., 2020) applied fluctuation correlation spectroscopy (FCS) as an independent means to estimate advection speed and diffusion coefficient. For this, they used a combination of RICS and TICS methods with correction for limited (masked) area (IVARICS). The authors of (Vartak et al., 2020) recognize that a major problem in the application of the method (described in Suppl. Materials of (Vartak et al., 2020)) is the macroscopic movement of the animal due to breathing and heartbeat. The authors state "*These movements typically occur in the range of 14ms -1s in a mouse. To avoid these artifacts in the spatial autocorrelation, we restricted the maximum pixel acquisition time to 8 µs, and the maximum image acquisition time to 250 µs. Since these movements occur on a slower timescale compared to molecular fluctuations, the high-pass filter applied to remove slowly moving structures effectively filters of the effect of animal movements*" (Vartak et al., 2020). Key statements are incorrect: 1) The image acquisition time was 316.5 msec (e.g. /tics/basal/mz/image 42.czi). There, a simple typo led to a wrong conclusion. A visual inspection of BC movements between sequential frames revealed that the animal movement makes TICS analysis not applicable.
2) The time between sequential lines on scanning images was ~1msec, much longer than the declared 250 µsec per frame. Also this could compromise RICS analysis and, indeed, in Appendix booklet (pp. 1–9 and 19-39) the RICS correlations randomly decay faster either on x- or on y- direction over the first 4 pixels (i.e. within BC size), dependent on BC orientation on the image. Given that the difference in time is 2 – 8 µsec for x and 1 – 4 msec for y, the correlation is significantly corrupted by tissue movement and is hardly conclusive. In contrast, the size of IBD is larger and tissue movements do not significantly affect the measurements. Consistent with this, the correlation in the x- direction always decays slower than in the y-direction, regardless of the IBD orientation on the image (Appendix booklet pp. 10 – 18). From this, we conclude that the IVARICS method is applicable to the IBD but not BC.

6. <u>Anisotropy of BC</u>. When quantifying flow, one has to consider that the BC network is anisotropic with some radial BC braches experiencing a steep hydrostatic pressure drop (see Fig. 6B of Meyer et al., 2017), but other more tangential BC branches experiencing no pressure drop and hence no flow. Most branches experience an intermediate case with slow flow. It is therefore expected that a „backbone" of radially oriented BC branches carries the main flow and short side branches even with stagnated flow diffusively deliver bile acids to a nearby backbone branch of the BC network. Transport is therefore advective over long distances. Diffusion may contribute and even dominate at the short scales of stagnant side branches. The IVARICS method neglects heterogeneities at the BC network level.

7. <u>Is absence of TCA effect conclusive?</u> Figure 6A of (Vartak et al., 2020) shows results for the uniform processing of the tracer signal. The quantification of the signal decay half-time in Fig. 6D of (Vartak et al., 2020) is interpreted by the authors of (Vartak et al., 2020) as absence of osmotic water influx and flow. This conclusion is based on the apparent large data variability upon TCA infusion but that was caused by a single outlier with three times the mean decay half-time of the other data points. Removing such outlier decreases the half-time distribution for TCA by almost one third, from ~650sec to ~450sec, faster than control.

8. <u>Arguments in favor of flow.</u> As diffusion is based on the Brownian motion of solutes in the BC, a diffusion-based flux would logically require a sufficiently steep gradient of concentration across the BC network with maximum near the central vein and minimum at IBD. In the absence of such a gradient the net diffusion-based flux would be zero. Localized photo-activation creates a sharp gradient of fluorescent tracer at the border around the illuminated volume. Such a sharp gradient leads to an artificially exaggerated diffusion-based flux of the photo-activated form of the tracer in comparison with its advection. However, there is also non-fluorescent tracer outside of the illuminated volume (diffusing into the activation volume) and the gradient of total tracer (fluorescent + non-fluorescent) remains flat and is not related to the fluorescent gradient. Hence, retrograde diffusion of the locally activated fraction of tracer is entirely expected, even for IBD with larger flow as shown in Fig.1C bottom row of (Vartak et al., 2020), and is no indication of the relation between diffusion and flow of total tracer or other more uniformly secreted molecules. The detailed theoretical considerations are provided here in the Appendix.

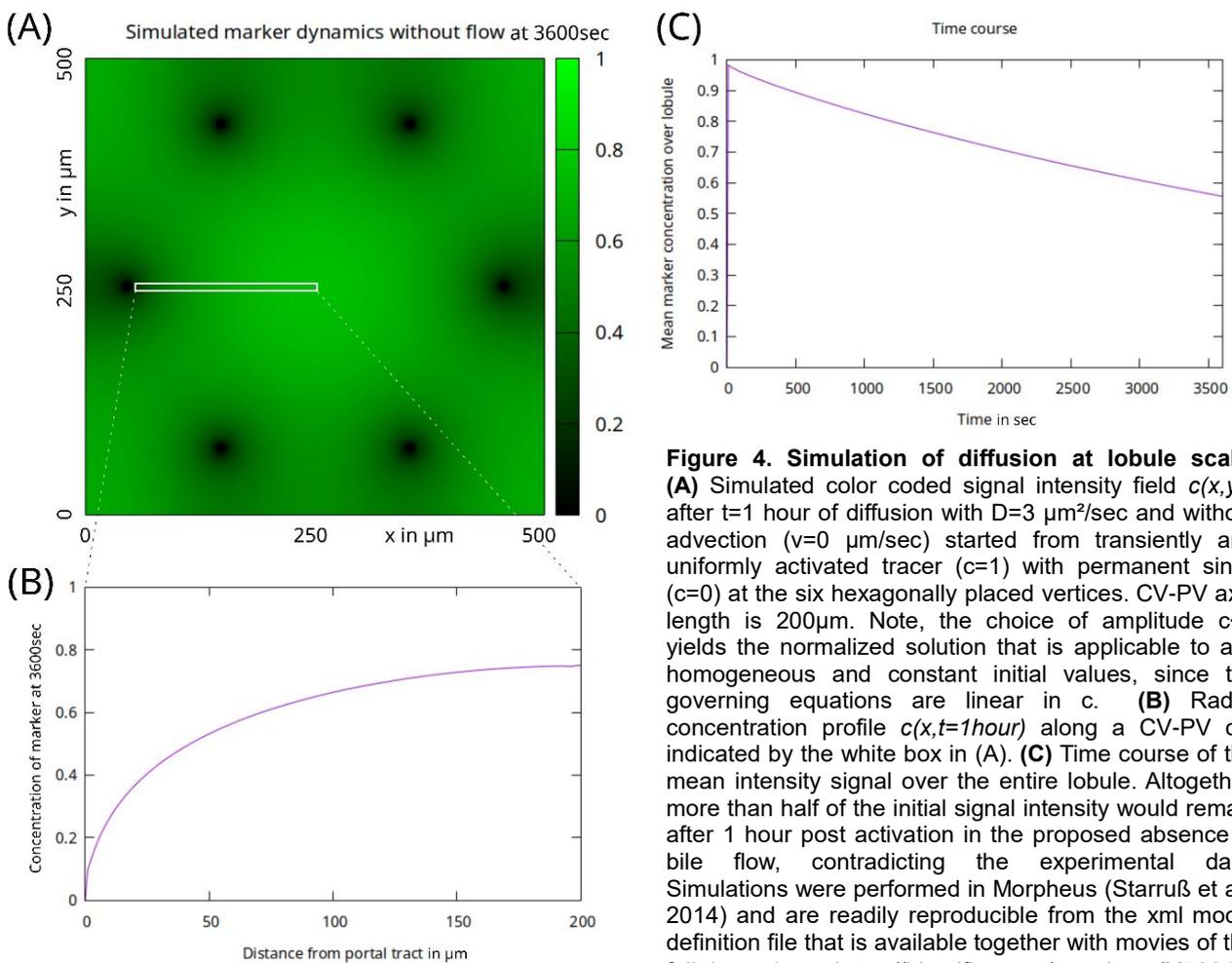

**Figure 4. Simulation of diffusion at lobule scale.** **(A)** Simulated color coded signal intensity field $c(x,y,t)$ after t=1 hour of diffusion with D=3 µm²/sec and without advection (v=0 µm/sec) started from transiently and uniformly activated tracer (c=1) with permanent sinks (c=0) at the six hexagonally placed vertices. CV-PV axis length is 200µm. Note, the choice of amplitude c~1 yields the normalized solution that is applicable to any homogeneous and constant initial values, since the governing equations are linear in c. **(B)** Radial concentration profile $c(x,t=1hour)$ along a CV-PV cut indicated by the white box in (A). **(C)** Time course of the mean intensity signal over the entire lobule. Altogether, more than half of the initial signal intensity would remain after 1 hour post activation in the proposed absence of bile flow, contradicting the experimental data. Simulations were performed in Morpheus (Starruß et al., 2014) and are readily reproducible from the xml model definition file that is available together with movies of the full dynamics at https://identifiers.org/morpheus/M7991.

The microscopy arguments (points 1 and 2 above) question the results of the published simulations in (Vartak et al., 2020), because these are based on a wrong assumption.

Simulations of the model equations (1,2) in (Vartak et al., 2020) would be needed at the lobule scale, but are not considered at lobule scale in (Vartak et al., 2020). Alternatively, if one simulates the model with uniform, isotropic diffusion (considering D=3 µm²/sec as in (Vartak et al., 2020)) and assumed absence of flow (v=0 µm/sec) at the lobule scale (200 µm CV-PV distance), the insufficiency of diffusion alone to clear the fluorescence signal within the observed time frame becomes evident (see Fig. 4). Moreover, this insufficient uniform diffusion throughout the lobule even overestimates the effect since diffusion along quasi-1D and non-straight branches of the BC network is less efficient. Figure 4 shows the concentration field after one hour of diffusion started from transiently and uniformly activated tracer (*c=1*) and with permanent sinks (*c=0*) at the six hexagonally placed vertices (representing CoH or IBD at portal triads, CV is in the center of the domain). Even in this simplified homogeneous space, diffusion alone cannot explain the experimentally observed clearance of initially uniformly activated tracer from the entire lobule with a mean decay half-time ~650sec (Fig. 6D in (Vartak et al., 2020)). Instead, with diffusion alone, more than half of the initial signal would remain in the central 2/3 of the lobule after 1 hour. In reality, the diffusion of small molecules would be even less efficient, as the basic principle of size exclusion chromatography teaches us. This contradiction readily vanishes when considering the advection process from bile flow (see Fig. 6 in Meyer et al., 2017).

In conclusion, text-books are not written in stone but serve as stimuli to prove or disprove what (we think) we know. New technologies, simulations and theories help us testing model predictions. However, the data obtained need to be interpreted within the limits of our technological means. We believe that the statement that bile transport in the BC is diffusion-dominated would require more rigorous evidence to disprove the text-book model.

**Acknowledgements**
We thank Sebastian Bundschuh and Jan Peychl of the MPI-CBG Light Microscopy Facility for discussions and estimations of photo-activation in relation to the PSF of Fig. 1. We also thank Michael Kücken for discussions. This work was financially supported by the Liver Systems Medicine program (LiSyM, grant #031L0038), funded by the German Federal Ministry of Research and Education (BMBF), the BMBF grant on "a systems microscopy approach to tissues and organ formation" (grant #031L0044), the European Research Council (grant #695646), and the Max Planck Society (MPG).

**Appendix**

Formalizing the transport problem allows us to quantify the relative importance of the two contributions, diffusion and advection, as follows. The transport of bile acids with concentration profile $c(\underline{x},t)$ at position $\underline{x}$ and time $t$ is governed by sources $g(\underline{x},t)$ through apical release from hepatocytes or fluorescence activation, advection with the steady state velocity profile $\underline{v}(\underline{x})$ and diffusion with coefficient $D$. In agreement with the first equation in the Suppl. Materials section „In silico modeling/2. Molecular transport simulation in bile/Theoretical considerations" of (Vartak et al., 2020), the transport equation reads

$$\frac{\partial c(\underline{x},t)}{\partial t} = g(\underline{x},t) - \underline{\nabla} \cdot \underline{j}(\underline{x},t) \tag{1}$$

with the net flux $\underline{j}$ of molecules determined by

$$\underline{j}(\underline{x},t) = -D\underline{\nabla}c(\underline{x},t) + \underline{v}(\underline{x},t)\, c(\underline{x},t) = -D\, c(\underline{x},t)\left[\frac{\underline{\nabla}c(\underline{x},t)}{c(\underline{x},t)} - \frac{\underline{v}(\underline{x},t)}{D}\right], \tag{2}$$

or likewise for the relative flux

$$\frac{\underline{j}(\underline{x},t)}{c(\underline{x},t)} = D\left[\frac{-\underline{\nabla}c(\underline{x},t)}{c(\underline{x},t)}\right] + \underline{v}(\underline{x},t). \tag{3}$$

Hence, the flux $\underline{j}$ has two contributions, the first from diffusion and the second from advection due to bile flow. While the second term always transports in the direction of flow ($\underline{j}\sim\underline{v}$), the first term can transport against flow if the relative spatial derivative of the concentration profile exceeds the (given) ratio of flow velocity and diffusion constant. Hence, any sharply localized elevations (likewise depressions) of concentration will show retrograde and qualitatively isotropic transport but such observations are no evidence for the absence of flow, rather a confirmation of heterogeneous signal perturbation.

Under physiological conditions with smooth $g(\underline{x},t)$, the diffusion process will level out any spatial heterogeneities such that relative gradients of $c$ become negligibly small over time and advection dominates at the physiologically relevant length and time scales.

As a consequence, local fluorescence activation as in FLAP is inherently highlighting diffusion effects and many of the results in Figs. 2 and 4 of (Vartak et al., 2020), notably periodic activation and retrograde signal spreading, are therefore expected in any case and no argument for or against flow in BC.

To quantify the relative importance of the two contributions, diffusion and advection, to signal transport, the dimensionless Péclet number (Pe=L*v/D) is used (see point 3). This Péclet number yields the ratio of the advective contribution over the diffusive contribution to the flux and L and v are the characteristic length and velocity scales. The magnitude of the transport flux j in equation (2) then scales as j~[1-Pe] times the diffusive flux contribution. Hence, a flux composition can be characterized as diffusion-dominated when the first term in equation (2), and equivalently the first term in [1-Pe], dominates over the second which can be quantitatively evaluated as Pe << 1. Conversely, a flux composition can be characterized as advection-dominated when the second term dominates over the first and therefore quantitatively, Pe >> 1.